\documentclass[11pt,epsf,letterpaper]{article}%
\pdfoutput=1
\usepackage{setspace}
\usepackage{color}
\usepackage{amsmath}
\usepackage{amsfonts}
\usepackage{verbatim}
\usepackage{amssymb}
\usepackage{graphicx}
\usepackage{amsmath,amssymb,calc}
\usepackage[bulletsep]{collref}
\usepackage{bbm}
\usepackage{setspace}
\usepackage{color}
\usepackage{amsmath}
\usepackage{amsfonts}
\usepackage{verbatim}
\usepackage{amssymb}
\usepackage{graphicx}
\usepackage{array}
\usepackage{caption}
\usepackage{subcaption}%
\providecommand{\U}[1]{\protect\rule{.1in}{.1in}}
\pdfoutput=1

\newcommand\be{\begin{equation}}
\newcommand\ee{\end{equation}}
\newcommand{\bead}{\begin{aligned}}
\newcommand{\eead}{\end{aligned}}

\newcommand{\bea}{\begin{eqnarray}}
\newcommand{\eea}{\end{eqnarray}}

\def\beq{\begin{equation}}
\def\eeq{\end{equation}}
\def\id{\protect{{1 \kern-.28em {\rm l}}}}

\def\unit{\relax{\rm 1\kern-.26em I}}

\begin{document}

\title{$SU(N)$ BPS Monopoles in $\mathcal{M}^{2}\times S^{2}$}
\author{Fabrizio Canfora$^{1}$, Gianni Tallarita$^{1,2}$\\$^{1}$\textit{Centro de Estudios Cient\'{\i}ficos (CECS), Casilla 1469,
Valdivia, Chile.}\\$^{2}$\textit{ Departamento de F\'{\i}sica, Universidad de Santiago de Chile,
Casilla 307, Santiago, Chile}\\{\small canfora@cecs.cl, tallarita@cecs.cl, gianni.tallarita@usach.cl}}
\maketitle

\begin{abstract}
We extend the investigation of BPS saturated t'Hooft-Polyakov monopoles in
$\mathcal{M}^{2}\times S^{2}$ to the general case of $SU(N)$ gauge symmetry.
This geometry causes the resulting $N-1$ coupled non-linear ordinary
differential equations for the $N-1$ monopole profiles to become autonomous.
One can also define a flat limit in which the curvature of the background
metric is arbitrarily small but the simplifications brought in by the geometry
remain. We prove analytically that non-trivial solutions in which the profiles
are not proportional can be found. Moreover, we construct numerical solutions
for $N=2,3$ and 4. The presence of the parameter $N$ allows one to take a
smooth large $N$ limit which greatly simplifies the treatment of the infinite
number of profile function equations. We show that, in this limit, the system
of infinitely many coupled ordinary differential equations for the monopole
profiles reduces to a single two-dimensional non-linear partial differential equation.

\end{abstract}

\section{Introduction}

The subject of 't Hooft-Polyakov monopoles \cite{'tHooft:1974qc}
\cite{Polyakov:1974ek} has been extensively studied in the literature, (see
for reviews \cite{Rossi:1982fq} \cite{manton} \cite{Tong:2005un}). The
fundamental role these objects are believed to play in modern theoretical
physics ranges between all branches of the subject (for a comprehensive review
see, for example, \cite{Rajantie:2012xg}). These solitonic objects have become
predominant characters of most modern-day non-Abelian gauge theories,
including supersymmetry \cite{susy2} and string theory (see for example
\cite{Polchinski:2003bq}) despite having never been seen in experiments. One
aspect of their description which is less studied is their intimate link with
the topology of the underlying space. This paper is an extension of our
previous work \cite{Canfora:2014jsa} which investigates solitonic
t'Hooft-Polyakov monopoles in a cylindrical topology with spherical
cross-sections (for related work see also \cite{Popov:2008wi} \cite{Popov:2008wh}). In our previous investigation we showed that for $SU(2)$
monopoles living in this topology all field profile equations become
autonomous (that is, they don't depend on the radial variable $r$ explicitly)
which allows one to treat them by analytical tools normally not available. In
this work we extend this set-up to the general case of $SU(N)$ gauge symmetry.
The main mathematical tool used to achieve this is the formalism of harmonic
maps which is reviewed in section 2. We will show that, as usually happens, by
introducing an arbitrary dependence on the parameter $N$, one can find find a
simplifying limit in which $N$ becomes large. The simplifications brought about by this limit are especially treatable in the harmonic map formalism and constitute the main reason for choosing this over other methods.\newline\indent The paper is
organized as follows: section 2 is devoted to introducing the system with its
Lagrangian and the aforementioned topology. This section also includes a
review of the harmonic map formalism used to describe $SU(N)$ monopoles. In section 3 we discuss in
detail the consequences of the large $N$ limit and in section 4 we find
solutions numerically for the cases $N=2,3,4$. Finally we provide some
conclusions in section 5. In the appendix we provide a rigorous mathematical proof  that the resulting
equations for field profiles always allow non-trivial solutions (given a
suitable bound on the shape of the cylinder).

\section{The System}

We begin this section by reviewing the general algorithm to obtain spherically
symmetric $SU(N)$ monopole solutions using rational maps of the Riemann sphere
into flag manifolds. We refer the reader to \cite{Ioannidou:1999iw}, on which
this review is based and from which we borrow our notation, for the relevant
mathematical details. \newline

The action of the $SU(N)$ Yang-Mills-Higgs system in four dimensional
space-time is
\begin{equation}
S_{\mathrm{YMH}}=\int d^{4}x\sqrt{-g}\;\mathrm{Tr}\left(  \frac{1}{4}F_{\mu
\nu}F^{\mu\nu}+\frac{1}{2}D_{\mu}\Phi D^{\mu}\Phi-\frac{\lambda^{2}}{8}\left(
\Phi\Phi-v^{2}\right)  ^{2}\right)  \ ,\label{action}%
\end{equation}
where the Planck constant, the speed of light and the gauge coupling constant
have been set to $1$. The remaining dimensionless coupling constant is
$\lambda$. In this equation, $\Phi$ is an $su(N)$ valued scalar field, the
covariant derivative is $D_{i}=\partial_{i}+\left[  A_{i},\right]  $ and
$F_{ij}=\partial_{i}A_{j}-\partial_{j}A_{i}+\left[  A_{i},A_{j}\right]  $. The
$SU(N)$ BPS monopoles are finite energy solutions of the BPS equation
\begin{equation}
D_{i}\Phi=-\frac{1}{2}\epsilon_{ijk}F^{jk},
\end{equation}
obtained by an energy minimization argument from eq.(\ref{action}) when
$\lambda=0$. In polar coordinates with $z=e^{i\varphi}\tan(\theta/2)$ the flat
space-time metric reduces to
\begin{equation}
ds^{2}=-dt^{2}+dr^{2}+\frac{4r^{2}}{(1+|z|^{2})^{2}}dzd\bar{z},
\end{equation}
and one can use the ansatz
\begin{equation}
\Phi=-iA_{r}=-\frac{i}{2}H^{-1}\partial_{r}H,\quad A_{z}=H^{-1}\partial
_{z}H,\quad A_{\bar{z}}=0,\label{ansatz}%
\end{equation}
for $H\in SL(N,\mathbb{C})$, to reduce the matrix system of BPS equations to
\begin{equation}
\partial_{r}\left(  H^{-1}\partial_{r}H\right)  +\frac{\left(  1+|z|^{2}%
\right)  ^{2}}{r^{2}}\partial_{\bar{z}}\left(  H^{-1}\partial_{z}H\right)
=0.\label{eom}%
\end{equation}
\indent The ansatz used in eq.(\ref{ansatz}) (which is the most natural
generalization of the 't Hooft-Polyakov hedgehog ansatz for $SU(2)$) describes
three-dimensional topological defects with non-Abelian magnetic charge given
by
\begin{equation}
Q_{M}=\int_{S_{R}}d^{2}S_{i}B_{i}^{a}\Phi^{a},\label{magnetic1}%
\end{equation}
where $S_{R}$ denotes a sphere centered around the monopole and $B_{i}%
^{a}=\frac{1}{2}\epsilon_{ijk}F_{jk}^{a}$ is the non-abelian magnetic field.
More generally, the dimensionality of a topological defect (see, for a
detailed explanation, \cite{manton} and \cite{susy}) is determined by the
homotopy class of the corresponding ansatz. In general, for systems living in a spacetime $\mathcal{M}$ and with a gauge symmetry $G$ broken down to a subgroup $H$ one can label the toplogical charges of solitonic solutions by the degree of maps of $S^n\in \mathcal{M}$ into the coset space $G/H$,
\be\label{class0}
\pi_n\left(G/H\right),
\ee
where $n$ varies according to $\mathcal{M}$ and which pattern of symmetry breaking is considered\footnote{For instance, in flat space, vortex-like
objects have a non-trivial first homotopy class $\pi_{1}(U(1))$ , monopole-like
objects (which are point-like in space) have a non-trivial second homotopy
class $\pi_{2}(SU(2)/U(1))$ while instanton-like solutions (which are point-like in
space-time) have a non-trivial third homotopy class $\pi_3(SU(N))$.}. For the case of $SU(N)$ monopoles considered here, we will be interested in the homotopy class
\be\label{class}
\pi_2\left(SU(N)/J\right),
\ee
where $J$ can vary between $U(N-1)$ and $U(1)^{N-1}$ corresponding to how the gauge symmetry is broken. The case $U(N-1)$ corresponds to minimal symmetry breaking whilst $U(1)^{N-1}$ is the case of maximum symmetry breaking and these depend on the vacuum expectation value of $\Phi$.
 These solitonic objects
are stable unrelated solutions of different energy minimization requirements,
however one of the interesting outcomes of the present analysis is that
different topological objects can actually be difficult to tell apart (at
least by looking at the equations of motion) when such topological objects are
analyzed within space-like regions with non-trivial topology. In particular,
we will show that the field equations for non-Abelian BPS monopoles
(possessing non-trivial second homotopy class as per eq.(\ref{class})) within the bounded tube-shaped
region defined in the next section (see \cite{Canfora:2014jsa}) are related to
the field equations of domain-wall objects in flat topology and with a similar homotopy class eq.(\ref{class0}) (see \cite{bolognesi}) through a
simple field redefinition.\newline

In order to find spherically symmetric monopole solutions to equation
(\ref{eom}) one can follow a simple algorithm: first, one needs the
spherically symmetric maps into $\mathbb{C}\mathbb{P}^{N-1}$ (see
\cite{Ioannidou:1999iw}) which are given by
\begin{equation}
f=\left(  f_{0},...,f_{j},...,f_{N-1}\right)  ^{t},\quad f_{j}=z^{j}%
\sqrt{\left(
\begin{array}
[c]{c}%
N-1\\
j
\end{array}
\right)  },
\end{equation}
where the expression in the square root denotes the standard binomial
coefficient. Then using the $\Delta$ operator defined as
\begin{equation}
\Delta f=\partial_{z}f-\frac{f\left(  f^{\dagger}\partial_{z}f\right)
}{|f|^{2}},
\end{equation}
and applying it iteratively $\Delta^{k}f=\Delta\left(  \Delta^{k-1}f\right)
$, $k=0,...,N-1$, one can construct a projector matrix
\begin{equation}
P_{k}=P\left(  \Delta^{k}f\right)  ,\quad P(f)=\frac{ff^{\dagger}}{|f|^{2}},
\end{equation}
satisfying $P^{\dagger}=P=P^{2}$ which is used to parametrize the general
$N\times N$ $SL(N,\mathbb{C})$ Hermitian matrix $H$ appearing in
eq.(\ref{eom}) by
\begin{equation}
H=\exp\left(  \sum_{i=0}^{N-2}g_{i}\left(  P_{i}-1/N\right)  \right)  ,
\label{hpara}%
\end{equation}
where $g_{i}$ are general profile functions which depend only on $r$ and the
$1/N$ factor denotes the identity matrix divided by $N$. Substituting
eq.(\ref{hpara}) into eq.(\ref{eom}) gives a general matrix of equations which
can be decoupled for each $\ddot{g}_{i}$ (here $\dot{}$ denotes
differentiation w.r.t $r$). More generally, following \cite{Ioannidou:1999mf}
one can find a convenient form of the resulting equations in terms of $N$ and
$l=0,...,N-2$, the index labelling the profile function $g_{l}$,
\begin{equation}
-\frac{2(l+1)}{N}\sum_{i=0}^{N-2}(i+1)\ddot{F}_{i}+2\sum_{k=0}^{l}\sum
_{i=k}^{N-2}\ddot{F}_{i}-\frac{2}{r^{2}}(l+1)(N-l-1)(\exp(F_{l})-1)=0,
\label{genem}%
\end{equation}
where $F_{l}=g_{l}-g_{l+1}$ and $F_{N-2}=g_{N-2}$. Decoupling these equations
gives $N-1$ equations for the profile functions, we include below the examples
for $SU(N)$ with $N=2,3$, which are respectively
\begin{equation}
\label{su2}\ddot{g_{0}}+\frac{2}{r^{2}}(1-e^{g_{0}})=0,
\end{equation}
and
\begin{equation}
-\ddot{g_{0}}+\frac{2}{r^{2}}\left(  e^{g_{0}-g_{1}}-1\right)  +\frac{2}%
{r^{2}}\left(  e^{g_{1}}-1\right)  =0, \label{todasimilar1}%
\end{equation}%
\begin{equation}
-\ddot{g_{1}}-\frac{2}{r^{2}}\left(  e^{g_{0}-g_{1}}-1\right)  +\frac{4}%
{r^{2}}\left(  e^{g_{1}}-1\right)  =0. \label{todasimilar2}%
\end{equation}

As shown in \cite{Ioannidou:1999iw}, these equations have analytic solutions
corresponding to magnetic monopoles. In the simplest case of $N=2$, the
solution is
\begin{equation}
g_{0}=2\log\left(  2r/\sinh2r\right)  , \label{sol}%
\end{equation}
which describes a single BPS saturated monopole, with energy equal to its
charge $Q_{M}$. In the case where $N>2$, the equations still allow for at
least one analytic solution in which the profile functions are chosen
proportional to each other. Indeed if, for example, we choose $N=3$ then there
exists the solution with $g_{0}=2g_{1}$. In general, one has a solution for
all field profiles proportional when
\begin{equation}
\label{prop}\frac{g_{i}}{N-(i+1)} = g
\end{equation}
with $g$ given by eq.(\ref{sol}) and $i=0,...,N-2$.

\subsection{$SU(N)$ monopoles on $\mathcal{M}^{2}\times S^{2}$}

In this section we discuss a simple geometrical modification of the above set
up and its consequences. We wish to consider the above system in
$\mathcal{M}^{2}\times S^{2}$ (or $\mathcal{M}^{1}\times S^{1}\times S^{2}$)
with metric
\begin{equation}
ds^{2}=-dt^{2}+dr^{2}+R_{0}^{2}(d\theta^{2}+\sin^{2}\theta\ d\phi
^{2})\ ,\ \ 0\leq r\leq L\ , \label{metric}%
\end{equation}%
\begin{equation}
0\leq\theta\leq\pi\ ,\ 0\leq\phi\leq2\pi\ , \label{range1}%
\end{equation}
where $L$ is a longitudinal length and $R_{0}$ is a constant with the
dimension of length related to the size of the transverse sections of this
topology. In complexified coordinates, those appropriate to this paper, the
metric reads
\begin{equation}
ds^{2}=-dt^{2}+dr^{2}+\frac{4R_{0}^{2}}{(1+|z|^{2})^{2}}dzd\bar{z}.
\label{metric}%
\end{equation}
This metric describes a tubular geometry with spherical caps as cross
sectional slices. The non-vanishing components of the Riemann tensor
$R_{\mu\nu\rho\sigma}$ of this space are proportional to $1/R_{0}^{2}$:%

\begin{equation}
R_{\mu\nu\rho\sigma}\sim\frac{1}{R_{0}^{2}}.
\end{equation}

This simple modification of the geometry leads to a dramatic simplification in
the resulting equations for the profile functions. As shown in
\cite{Canfora:2014jsa}, for the case of $SU(2)$, the resulting energy
minimization equations for the monopole profile functions are autonomous, that
is they don't involve any explicit powers of $r$. Below we find that this also
happens for the $SU(N)$ case. As shown later in the paper, one can easily
define a flat limit in which the curvature of the background metric is as
small as one wants keeping, at the same time, all the simplifications brought
in by the background metric in eq. (\ref{metric}) (two different ways to
achieve the flat limit will be described in the following sections). Moreover,
the present formalism introduces the possibility to define a large $N$ limit.
We will consider this in detail in a further section.

Equation (\ref{eom}) when analyzed in the background metric given by
eq.(\ref{metric}) is modified to
\begin{equation}
\partial_{r}\left(  H^{-1}\partial_{r}H\right)  +\frac{\left(  1+|z|^{2}%
\right)  ^{2}}{R_{0}^{2}}\partial_{\bar{z}}\left(  H^{-1}\partial_{z}H\right)
=0,
\end{equation}
and equation (\ref{genem}) becomes,
\begin{equation}
-\frac{2(l+1)}{N}\sum_{i=0}^{N-2}(i+1)\ddot{F}_{i}+2\sum_{k=0}^{l}\sum
_{i=k}^{N-2}\ddot{F}_{i}-\frac{2\left(  l+1\right)  }{R_{0}^{2}}%
(N-l-1)(\exp(F_{l})-1)=0, \label{set}%
\end{equation}
where, as in the previous section, $F_{l}=g_{l}-g_{l+1}$. The equations now
become autonomous. In line with the discussion around eq.(\ref{prop}) we find
that for $N>2$ there exist analytical solutions in which all the profile
functions are proportional. However, in the following we will show both
numerically and analytically that non-trivial solutions in which the profiles
are not equal (and hence do not correspond to trivial embeddings of $SU(2)$
into $SU(N)$) also exist. \newline

Let us take a moment here to connect the previous results (and the more
standard monopole notation) to the current notation. For the simplest case of
$SU(2)$ the equation reduces to
\begin{equation}
\ddot{g_{0}}+\frac{2}{R_{0}^{2}}(1-e^{g_{0}})=0. \label{su2}%
\end{equation}
It was shown in \cite{Canfora:2014jsa} that for $N=2$ using standard polar
coordinates $(r,\theta,\phi)$ and an ansatz of the form
\begin{equation}
A_{\mu}=\left(  k\left(  r\right)  -1\right)  U^{-1}\partial_{\mu}%
U\ ,\ \ \Psi=\psi(r)U\ , \label{hedge1}%
\end{equation}%
\begin{equation}
U=\widehat{n}^{i}t_{i}\ ,\ \ U^{-1}=U\ , \label{standard1}%
\end{equation}%
\begin{equation}
\widehat{n}^{1}=\sin\theta\cos\phi\ ,\ \ \ \widehat{n}^{2}=\sin\theta\sin
\phi\ ,\ \ \ \widehat{n}^{3}=\cos\theta\ , \label{standard2}%
\end{equation}
where the $t^{i}$ are the standard Pauli matrices, the BPS equations reduced
to
\begin{equation}
\partial_{r}\psi+\frac{1-k^{2}}{R_{0}^{2}}=0, \label{BPSv2}%
\end{equation}
\begin{equation}
\label{BPSv3}\partial_{r} k -k\psi=0,
\end{equation}
for which a general solution was proposed of the form
\begin{equation}
\label{rel1}k=\exp(u)\ ,
\end{equation}%
\begin{equation}
\label{rel2}\psi=\partial_{r}u\ ,
\end{equation}

where the function $u(r)$ is the inverse of the following integral%
\begin{equation}
\int_{u(0)}^{u(r)}\left[  2\left(  I_{0}+\frac{\exp\left(  2z\right)
-z}{R_{0}^{2}}\right)  \right]  ^{-1/2}dz=\pm r\ , \label{integral}%
\end{equation}
with $I_{0}$ an integration constant. This is consistent with the above
construction, as one indeed expects, upon the identification $u=g_{0}/2$ where
eqs.(\ref{BPSv2}) and (\ref{BPSv3}) reduce to eq.(\ref{su2}). \newline

The autonomous set of equations (\ref{set}), of which eq.(\ref{su2}) is both
the $N=2$ and the degenerate equation for proportional fields representative,
has some interesting properties which we discuss throughout the paper. One of
which is that it coincides with the equation for a domain wall separating the
Higgs and Coulomb phases in the Abelian - Higgs model\footnote{We thank S. Bolognesi for pointing this out to us.}. In \cite{bolognesi},
the equation for the scalar field describing the domain wall was found to be
\begin{equation}
(\log\phi)^{\prime\prime2}=e^{2}\left(  \phi^{2}-v^{2}\right)  ,\label{wall}%
\end{equation}
where $e$ is the $U(1)$ gauge coupling and $v^{2}$ is the constant appearing
in the quartic potential (similar to that appearing in eq.(\ref{action})). The
above equation is a particular case of the well known Taubes equation
\cite{taubes}. Upon identifying $\phi=v\exp(g_{0}/2)$ and $v^{2}e^{2}%
=1/R_{0}^{2}$ we see that eq.(\ref{wall}) becomes eq.(\ref{su2}). What is
intriguing about this observation is that topological objects which are quite
different (possessing different non-trivial homotopy classes) can be described
by identical solutions if they are constrained to live in space-time regions
with non-trivial topology. In fact, since there is always a solution of the
$N$ equations in which the field profiles are proportional as in
eq.(\ref{prop}), this relation can be trivially extended to monopoles of
topological charge $N$.

\section{The large $N$ limit}

The large $N$ limit introduced in Yang-Mills theory in \cite{largeN1} (see
also \cite{largeN2} and \cite{LargeN3}; for two detailed reviews see
\cite{Nreview}) is a very powerful tool to analyze non-perturbative features
in gauge theories (such as confinement, bound states and so on). The
non-trivial scaling behavior of physical quantities within the large $N$
expansion arises from the fact that Feynman diagrams have different weights
depending on the topology of the surfaces they can be drawn on (once $N$ is
considered as a large number). Thus, this non-trivial behavior is purely
\textquotedblleft quantum" in nature and one would expect that no non-trivial
large $N$ behavior should be found when analyzing classical BPS equations (as
we are doing in the present paper). In fact, the present formalism shows that
a non-trivial scaling with $N$ emerges already at the level of the BPS field
equations eq.(\ref{set}). Naively taking $N$ very large means increasing the
number of non-linear coupled equations, or equivalently the number of monopole
profiles, which seems like an un-neccesary complication.

Fortunately, there is another dramatic simplification in this limit which
makes this problem treatable. The key issue is that in this limit, through an appropriate ansatz of the interpolating discrete function, one can
replace the discrete label $l=0,...,N-2$ in eq.(\ref{set}) by a
\textit{continuous label}. The main reason is that in the limit in which
$N\rightarrow\infty$, where the range of values of $l$ becomes infinite, each
discrete jump in its value (which is just $1$ of course) becomes infinitesimal
compared to the range. In other words, were we to rescale each step $\Delta
l=1$ of the complete range by $N$, this would become infinitesimal in this
limit. \newline

Let us consider this simplification in detail. First we take eq.(\ref{set})
and replace the discrete labels $F_{i}$ by a discrete function $F(i)$ (we omit
the dependence on the radial variable for simplicity). This is simply a
renaming which is useful to understand the subsequent manipulations. Clearly,
since $i$ is integer, this is a function on $\mathbb{Z}$. Then we multiply
eq.(\ref{set}) by $1/N^{2}$ and define the quantity
\begin{equation}
x=l/N.\label{xeq}%
\end{equation}
Then, in the limit in which $N\rightarrow\infty$, the discrete label $l$ can
be replaced by the continuous variable $x\in\mathbb{R}_{[0,1]}$. Consequently
we can replace the discrete function $F(l)$ living in $\mathbb{Z}$ by the
continuous function $F(x)$ living in $\mathbb{R}_{[0,1]}$. The equation then
becomes
\begin{equation}
-2\left(  x+\frac{1}{N}\right)  \sum_{i=0}^{N-2}\frac{(i+1)}{N}\frac{1}%
{N}\ddot{F}(i)+2\sum_{k=0}^{l}\frac{1}{N}\sum_{i=k}^{N-2}\frac{1}{N}\ddot
{F}(i)-\frac{2}{R_{0}^{2}}\left(  x+\frac{1}{N}\right)  \left(  1-(x+\frac
{1}{N})\right)  \left(  \exp(F(Nx)-1)\right)  =0.
\end{equation}
Now let us manipulate the sums, define
\begin{equation}
i/N=y,\quad z=k/N,\quad p=i/N,
\end{equation}
then simple manipulations lead to
\begin{align}
-2\left(  x+\frac{1}{N}\right)  \sum_{y=0}^{1-1/N}\left(  y+1/N\right)
\frac{1}{N}\ddot{F}(Ny)+2\sum_{z=0}^{x}\frac{1}{N}\sum_{p=z}^{1-2/N}\frac
{1}{N}\ddot{F}(Np)\nonumber\\
\quad-\frac{2}{R_{0}^{2}}\left(  x+\frac{1}{N}\right)  \left(  1-(x+\frac
{1}{N})\right)  \left(  \exp(F(Nx)-1)\right)   &  =0.\label{spaccimma}%
\end{align}
Finally, taking the large $N\rightarrow\infty$ limit, in which we can safely
replace the sums by integrals%
\[
\sum_{y=0}^{1-1/N}\frac{1}{N}\left[  G(y)\right]  \underset{N\rightarrow
\infty}{\rightarrow}\int_{0}^{1}dy\;G(y)\ ,
\]
since, from eq.(\ref{xeq}), $1/N$ plays the role of $``dx"$ in the
mathematical definition of Riemann and Lebesgue integrals, expression
eq.(\ref{spaccimma}) greatly simplifies and becomes
\begin{equation}
-2x\int_{0}^{1}y\left[  \frac{\partial^{2}}{\partial r^{2}}G(y,r)\right]
dy+2\int_{0}^{x}\int_{z}^{1}\left[  \frac{\partial^{2}}{\partial r^{2}%
}G(y,r)\right]  dydz-\frac{2x}{R_{0}^{2}}(1-x)\left(  \exp(G(x,r))-1\right)
=0,\label{int}%
\end{equation}
where we defined $G(x,r)=F(Nx,r)$: $G(x,r)$\ 
will be denoted as the \textit{complete profile function} since it encodes
information about all the elementary profiles $F_{i}(r)$ at the same time.

From here throughout the rest of the paper (unless specified) we switch to
dimensionless units
\begin{equation}
\rho=vr,\quad\tilde{R_{0}}=vR_{0},\quad\tilde{E}=E/v,\label{rescaling}%
\end{equation}
where $v$ is the parameter appearing in the potential with dimensions of mass
and $E$ is the energy of the solution. Then eq.(\ref{int}) becomes
\begin{equation}
-2x\int_{0}^{1}y\left[  \frac{\partial^{2}}{\partial\rho^{2}}G(y,\rho)\right]
dy+2\int_{0}^{x}\int_{z}^{1}\left[  \frac{\partial^{2}}{\partial\rho^{2}%
}G(y,\rho)\right]  dydz-\frac{2x}{\tilde{R_{0}}^{2}}(1-x)\left(  \exp
(G(x,\rho))-1\right)  =0\ .
\end{equation}
Taking two derivatives with respect to $x$ gives single non-linear elliptic
partial differential equation for the complete profile function $G(x,\rho)$
which reads
\begin{equation}
\frac{\partial^{2}}{\partial\rho^{2}}G+\frac{\partial^{2}}{\partial x^{2}%
}\left[  \frac{x(1-x)}{\tilde{R_{0}}^{2}}\left(  \exp(G(x,\rho))-1\right)
\right]  =0\label{largeN}%
\end{equation}
Solving this equation with the boundary conditions equal to the profile
functions in the radial direction is expected to yield a function which
interpolates smoothly between the monopole profiles. This is a non-trivial
numerical task which we defer to a later publication. The main important
advantage of eq. (\ref{largeN}) over the original system in eq.(\ref{set}) is
that instead of having, in the large $N$ limit, \textit{infinitely many
coupled equations for infinitely many unknowns} one has \textit{just one
non-linear partial differential equation for the complete profile function}
$G\left(  x,\rho\right)  $. Eq. (\ref{largeN}) is well suited for
numerical analysis and one could also easily apply the techniques described in
the previous section to prove the existence of solutions with suitable
properties

It is worth emphasizing here that the only assumption used to derive eq.(\ref{int}) starting from the original system in eq.(\ref{set}) is that there
exists a smooth interpolating function $F(x)$ in which the group theoretical label
$\frac{l}{N}$ becomes a continuous variable $x$. A necessary condition is the existence of a large N limit of the
field theory. Although the proof of this result is not available yet, there is a huge amount
of evidence in the literature (see \cite{Nreview}\ and references therein)
that it is actually possible to define a smooth large $N$ limit in many
different areas (from gauge theories to matrix model and so on). Whether the profile functions $F_{l}(r)$ as functions of
the discrete label $l$ follow a pattern regular enough to be interpolated by a
continuous (smooth) function depends on the $g_i$ solutions at large $N$. Therefore, as used in this paper, this assumption translates to an ansatz on the solutions we consider in the large $N$ limit which, in detail, involves the function $F_{l}(r)$ to be doubly differentiable. This ansatz leads directly to eq.(\ref{largeN}). Remarkably the above argument can also be directly extended to the case in which one starts off with a flat topology and considers standard $SU(N)$ BPS monopoles in flat space.

We will now discuss two different ways to consider the flat limit
in the background geometry in eq.(\ref{metric}) in which $\tilde{R_{0}}%
^{2}\rightarrow\infty$.

\subsection{The flat limit as a perturbative limit}

Another important advantage of eq.(\ref{largeN}) over the original system
eq.(\ref{set}) is that, in the large $N$ limit, it discloses the role of the
adimensional curvature parameter $\tilde{R_{0}}^{-2}$ as coupling constant of
the master equation given by eq.(\ref{largeN}) in such a way that the flat
limit corresponds to a ``weak field" limit.

Indeed, from eq.(\ref{largeN}) it is clear that in the large $\tilde{R_{0}%
}^{2}$ limit (in which the curvature vanishes) the equation for $G\left(
x,\rho\right)  $ becomes just a linear partial differential equation while the
limit in which $\tilde{R_{0}}^{2}$ is small (so that the curvature is large)
the non-linear effects become strong as well. Thus, this formalism provides
one with a clear perturbative scheme in which the equation for the complete
profile function $G\left(  x,\rho\right)  $ becomes linear in the flat limit
in such a way that the curvature parameter $1/\tilde{R_{0}}^{2}$ plays the
role of a coupling constant. \newline

In fact, if one would put the monopoles on a flat geometry from the very
beginning the corresponding system would be eq.(\ref{genem}) and such a system
does not admit any obvious perturbative scheme in the large $N$ limit although
one would arrive at an equation similar to eq.(\ref{largeN}). Indeed,
following similar steps to the previous section, if one would start from
eq.(\ref{genem}) assuming in the large $N$ limit a continuous dependence on
the group label $l$, then one would arrive at the following large $N$ limit
for the complete profile function $G_{F}(x,\rho)$:%
\begin{equation}
\frac{\partial^{2}}{\partial\rho^{2}}G_{F}+\frac{\partial^{2}}{\partial x^{2}%
}\left[  \frac{x(1-x)}{\rho^{2}}\left(  \exp(G_{F}(x,\rho))-1\right)  \right]
=0\ , \label{flatN}%
\end{equation}
where the label $F$ has been added to emphasize that $G_{F}(x,\rho)$ is the
complete profile function of the system of monopoles described by
eq.(\ref{genem}) which live, from the very beginning, on a flat metric. The
difference between eq.(\ref{largeN}) and eq.(\ref{flatN}) is then apparent: in
the former equation $\tilde{R_{0}}^{-2}$ plays the role of coupling constant
allowing a perturbative analysis of the equation while in the latter equation
one cannot do this since $\tilde{R_{0}}^{-2}$ has been replaced by $\rho^{-2}$
where $\rho$ is one of the independent variables of the equation.

\subsection{The flat limit as a geometrical bound}

An alternative way to define a flat limit for the background metric
corresponds to rescaling the original longitudinal variable in eq.(\ref{set})
such that%
\begin{equation}
\tilde{\rho}=r/R_{0}\ . \label{flat1.1}%
\end{equation}
Since the length of the tube-shaped region in which the monopoles are living
is $L$, the above rescaling is equivalent to considering a tube of
adimensional length%
\begin{equation}
\widetilde{L}=\frac{L}{R_{0}}\ . \label{flat1.2}%
\end{equation}
Then, one can take the flat limit taking $R_{0}$ and $L$ simultaneously large
in such a way that $\widetilde{L}$ stays finite:%
\begin{equation}
\left.  R_{0}\rightarrow\infty\ ,\ \ L\rightarrow\infty\ \right\vert
\ \ \widetilde{L}=const\neq0,\ \infty\ . \label{flat1.3}%
\end{equation}
In this limit, the master equation eq.(\ref{largeN}) for the complete profile
function simply becomes%
\begin{equation}
\frac{\partial^{2}}{\partial\tilde{\rho}^{2}}G+\frac{\partial^{2}}{\partial
x^{2}}\left[  x(1-x)\left(  \exp(G(x,\tilde{\rho}))-1\right)  \right]  =0\ .
\label{flat1.4}%
\end{equation}
As shown in the appendix it is important to note however that, in order to be sure that when
considering the flat limit in eqs.(\ref{flat1.1}), (\ref{flat1.2}) and
(\ref{flat1.3}) non-trivial solutions always exist, one should take the limit
in such a way that the inequalities\footnote{In the generic $SU(N)$ case, the
application of the Schauder theorem would give very similar inequalities.} in
eqs.(\ref{ban10.5}) and (\ref{ban11.5}) are never violated. This fact can be
interpreted as a sort of bound on the shape of the cylinder which cannot be
too ``slim" since $\widetilde{L}\gg1$ would violate eqs.(\ref{ban10.5}) and
(\ref{ban11.5}).

\section{Numerical Solutions}

In this section we provide numerical solutions for the profile functions in
the specific cases of $N=2,3$ and $4$. For every value of $\tilde{R_{0}}$
there exist a unique solution for the profile functions with an integer value
of $Q_{M}$. The numerical strategy is therefore the following: we provide
boundary conditions for the profile functions and vary $\tilde{R_{0}}$ until a
solution with the desired topological charge is found. This guarantees that
the solution obtained is a solution (for a given $\tilde{R_{0}}$) which
represents a BPS monopole with topological charge $Q_{M}$. The finite length
cylindrical topology is implemented numerically by imposing the boundary
conditions at a finite cutoff. The numerical procedure is a second-order
finite difference procedure with accuracy $\mathcal{O}(10^{-4})$.

\subsection{$N=2$}

Let us proceed to solve equation (\ref{su2}) numerically. We fix $v L =50$. We
fix the boundary conditions on the profile function to be
\begin{equation}
g_{0}(0)=1, \quad g_{0}(50)=0,
\end{equation}
and we find that at $\tilde{R_{0}}= 1.45$ the energy reads
\begin{equation}
\tilde{E} = \frac{Q_{M}}{4\pi} = \left[  \left(  e^{g_{0}}-1\right)
\frac{g_{0}^{\prime}}{2}\right]  \bigg|_{\rho=0}^{\rho=50} = 1.000,
\end{equation}
which is the expected one-monopole solution. This solution is shown in figure 1.

\begin{figure}[ptb]
\centering
\includegraphics[width=0.6\linewidth]{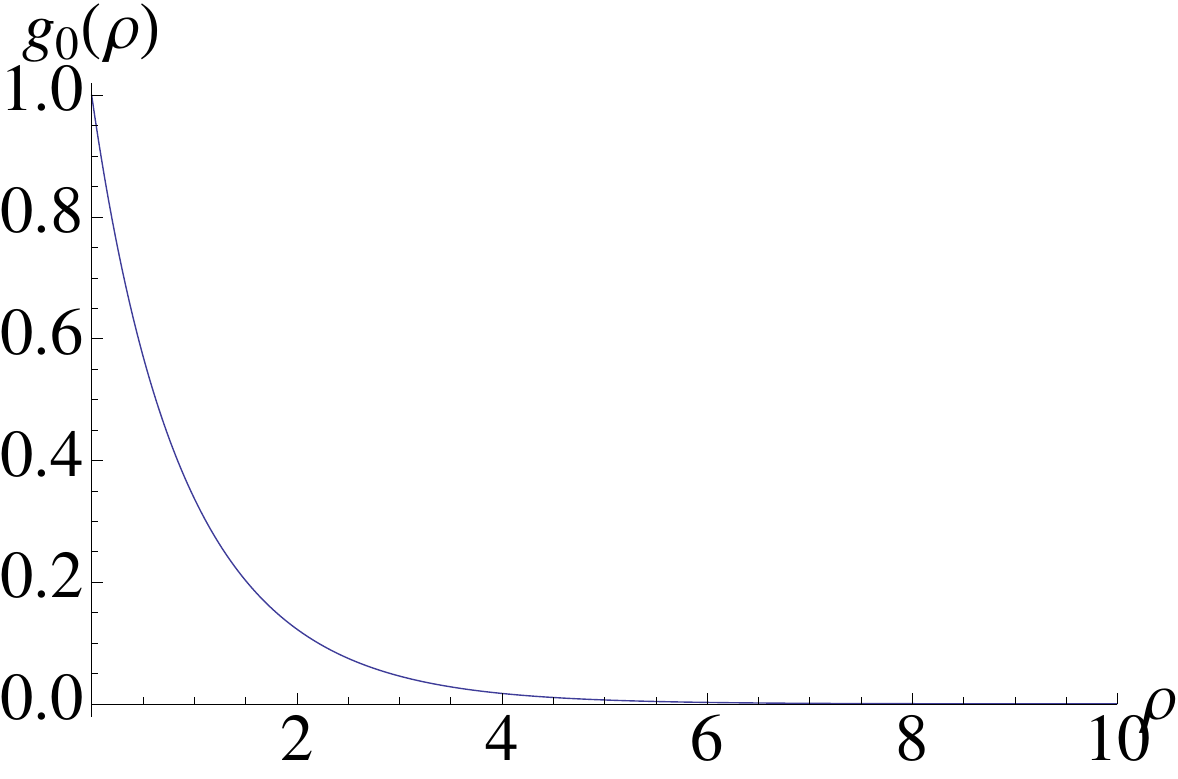} \caption{BPS monopole
profile function for $SU(2)$. The plot corresponds to the numerical value (in
appropriate units of $v$) $\tilde{R_{0}} = 1.45$.}%
\label{fig1}%
\end{figure}

For this value of $\tilde{R_{0}}$ one can use the topology of the system to
find the solution corresponding to placing the monopole at the other end of
the cylinder. This solution is found by imposing the conditions
\begin{equation}
g_{0}(0)=0,\; g_{0}(50)=1,
\end{equation}
and is shown in figure 2. For this solution, which is just an inversion of the
previous solution about the center of the cylinder, the energy is the same, as
expected. Note that this is not an anti-monopole solution as the topological
charge is the same. \newline

At this point a comment is in order regarding the $SU(2)$ solutions of
\cite{Canfora:2014jsa}. Below eq.(\ref{integral}) we pointed out that there is
a direct map between the profile function $g_{0}$ and the solutions obtained
without using the harmonic map formalism. Namely we showed that using
$u=g_{0}/2$ one can map the two first order BPS equations into the profile
equation for $g_{0}$. However, the solution for $g_{0}$ presented in figure 1,
when translated to the standard Higgs and gauge fields $\psi=g_{0}^{\prime}/2$
and $k=\exp(g_{0}/2)$ does not reproduce the solution found in
\cite{Canfora:2014jsa} even though both solutions have the same topological
charge (the reader may argue that the values of $\tilde{R_{0}}$ used for both
plots are not the same, but this comment applies also when these values are
made equal). This is however expected, in order for both solutions to match,
since $k(0)=0$, one would have to provide a singular boundary condition for
the profile function $g_{0}(0)=-\infty$, which is numerically impossible. The
non-uniqueness of the solution with given charge is made possible because in
this topology one does not require that the profile function be regular at the
radial origin, i.e. $g_{0}^{\prime}=0$ precisely because (unlike what happens
in flat metric in spherical coordinates) there is no preferred origin.

\begin{figure}[ptb]
\centering
\includegraphics[width=0.6\linewidth]{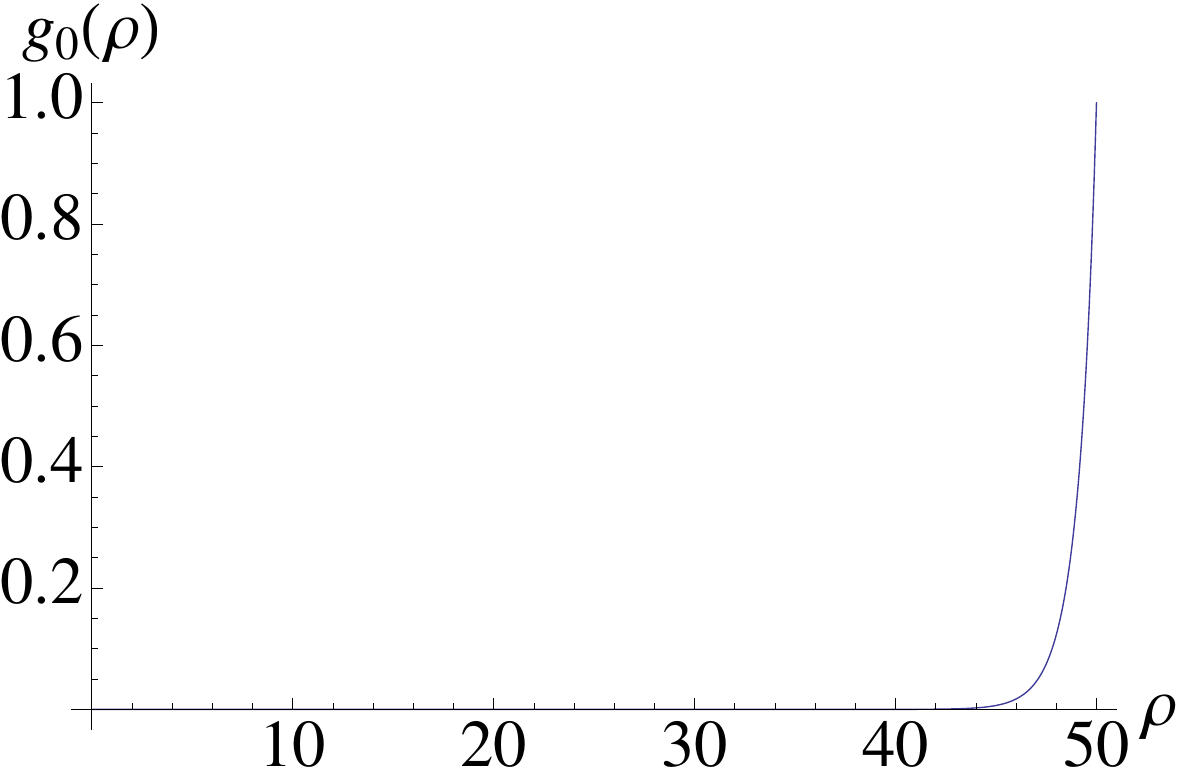} \caption{BPS monopole
profile function for $SU(2)$ with boundary conditions at other end of the
cylinder. The plot corresponds to the numerical value (in appropriate units of
$v$) $\tilde{R_{0}} = 1.45$.}%
\label{fig1}%
\end{figure}

\subsection{$N=3$}

In this section we wish to solve
\begin{equation}
-g^{\prime\prime}_{0}+\frac{2}{\tilde{R_{0}}^{2}}\left(  e^{g_{0}-g_{1}%
}-1\right)  +\frac{2}{\tilde{R_{0}}^{2}}\left(  e^{g_{1}}-1\right)  =0
\label{N3.1}%
\end{equation}%
\begin{equation}
-g^{\prime\prime}_{1}-\frac{2}{\tilde{R_{0}}^{2}}\left(  e^{g_{0}-g_{1}%
}-1\right)  +\frac{4}{\tilde{R_{0}}^{2}}\left(  e^{g_{1}}-1\right)  =0.
\label{N3.2}%
\end{equation}
where $^{\prime}$ denote differentiation w.r.t $\rho$. In Figure 2 we show the
solution corresponding to the case where the profile functions are
proportional. This is easily obtained by demanding boundary conditions of the
form
\begin{equation}
g_{0}(0)=1,\quad g_{1}(0)=0.5
\end{equation}%
\begin{equation}
g_{0}(50)=g_{1}(50)=0.
\end{equation}

The general energy equation reads
\begin{equation}
\label{su3en}\tilde{E} = \frac{Q_{M}}{4\pi} = e^{-g_{1}}\left[  \left(
e^{g_{0}}-e^{g_{1}}\right)  \dot{g_{0}}-\left(  e^{g_{0}}-e^{2g_{1}}\right)
g^{\prime}_{1}\right]  \bigg|_{\rho=0}^{\rho=50}%
\end{equation}
For the solution in Figure 3, by construction we find numerically%
\begin{equation}
\tilde{E}=1.0006.
\end{equation}
We can however also look for solutions which are not proportional. These are
more interesting solutions from the numerical point of view as they are full
solutions of the coupled equations rather than the reduction of all of these
to one equation for a single profile function. \newline Using the topology of
the system we may also find solutions which have non-vanishing boundary
conditions at both ends of the cylinder and therefore not proportional. In
this case we seek solutions with boundary conditions of the form
\begin{equation}
g_{0}(0)=1,\quad g_{1}^{\prime}(0)=0,
\end{equation}%
\begin{equation}
g_{0}^{\prime}(50)=0,\quad g_{1}(50)=1,
\end{equation}
these are shown in Figure 4 for a very specific choice of $\tilde{R_{0}}$. The
energy for this solution, calculated numerically using eq.(\ref{su3en}) is
\begin{equation}
\tilde{E}=2.001,
\end{equation}
and hence it is tempting to identify this solution as a monopole-anti-monopole
state, with each species of monopole living at each end of the cylinder.
However once again the boundary conditions lead to monopole charges which are
not consistent with this picture.

\begin{figure}[ptb]
\centering
\includegraphics[width=0.7\linewidth]{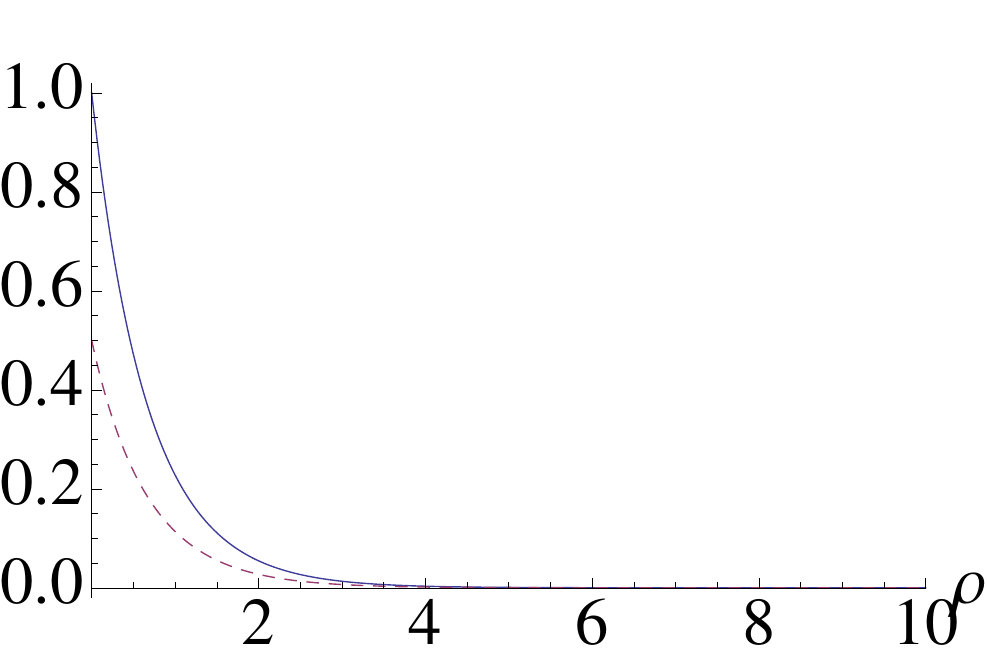} \caption{Proportional
monopole profile functions for $SU(3)$. The plots correspond to the numerical
value $\tilde{R_{0}} =1$. For these plots $g_{0} = 2g_{1}$. $g_{0}$ is the
solid line whilst $g_{1}$ is the dashed line.}%
\label{fig2}%
\end{figure}

\begin{figure}[ptb]
\centering
\includegraphics[width=0.7\linewidth]{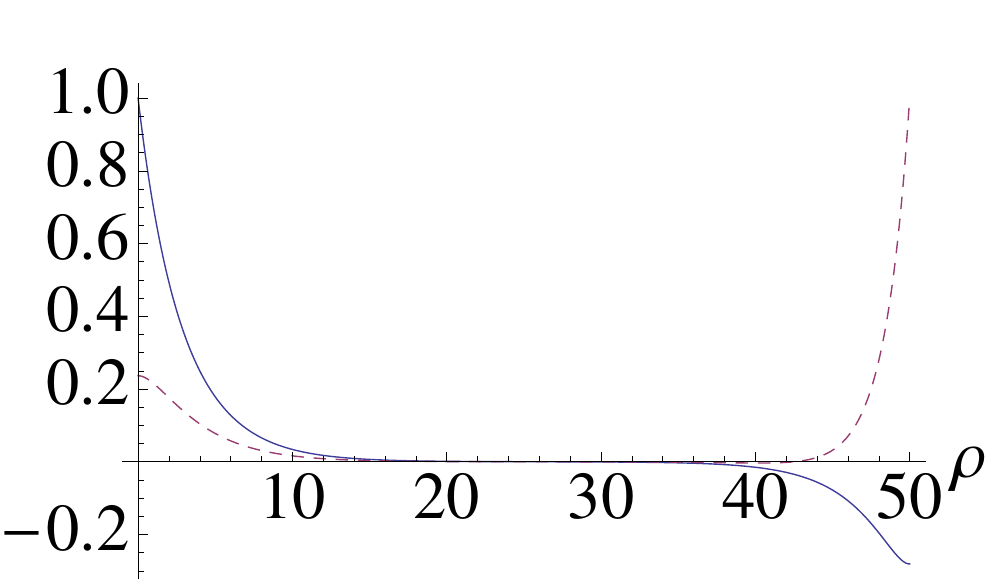} \caption{A solution with
non-vanishing boundary conditions at both ends of the cylinder at
$\tilde{R_{0}} = 4.28$. Dashed line is $g_{1}$, solid line is $g_{0}$.}%
\label{figmix}%
\end{figure}

\subsection{$N=4$}

In this case we wish to solve the system of equations
\begin{equation}
-g^{\prime\prime}_{0}+\frac{3}{\tilde{R_{0}}^{2}}\left(  e^{g_{0}-g_{1}%
}-1\right)  +\frac{3}{\tilde{R_{0}}^{2}}\left(  e^{g_{2}}-1\right)  =0,
\label{N4.1}%
\end{equation}%
\begin{equation}
-g_{1}^{\prime\prime}-\frac{3}{\tilde{R_{0}}^{2}}\left(  e^{g_{0}-g_{1}%
}-1\right)  +\frac{4}{\tilde{R_{0}}^{2}}\left(  e^{g_{1}-g_{2}}-1\right)
+\frac{3}{\tilde{R_{0}}^{2}}\left(  e^{g_{2}}-1\right)  =0, \label{N4.2}%
\end{equation}%
\begin{equation}
-g_{2}^{\prime\prime}-\frac{4}{\tilde{R_{0}}^{2}}\left(  e^{g_{1}-g_{2}%
}-1\right)  +\frac{6}{\tilde{R_{0}}^{2}}\left(  e^{g_{2}}-1\right)  =0,
\label{N4.3}%
\end{equation}
and the energy evaluates to
\begin{equation}
\tilde{E}=\frac{Q_{M}}{4\pi}=e^{-g_{1}-g_{2}}\left[  3e^{g_{0}+g_{2}}\left(
g_{0}^{\prime}-g_{1}^{\prime}\right)  +4e^{2g_{1}}\left(  g_{1}^{\prime}%
-g_{2}^{\prime}\right)  +e^{g_{1}+g_{2}}\left(  -3g_{0}^{\prime}-g_{1}%
^{\prime}+\left(  1+3e^{g_{2}}\right)  g_{2}^{\prime}\right)  \right]
\bigg|_{\rho=0}^{\rho=50}.
\end{equation}

We look for the analogous two-monopole solution of the previous section.
Therefore we impose the boundary conditions
\begin{equation}
g_{0}(0)=1, \;g_{1}^{\prime}(0)=0, \;g_{2}^{\prime}(0)=0,
\end{equation}
\begin{equation}
g_{0}^{\prime}(50)=0,\; g_{1}(50)=1, \; g_{2}^{\prime}(50)=0.
\end{equation}
The corresponding solution shown in Figure 5 is found to have $\tilde
{E}=2.000$ at $\tilde{R_{0}}=5.9$. If one calculates the topological charge at
each end of the cylinder one once again does not find equal contributions.
\newline

In each case, it appears that the profile functions concentrate around the
end-points of the cylindrical topology and want to vanish in the intermediate
region. \begin{figure}[ptb]
\centering
\includegraphics[width=0.6\linewidth]{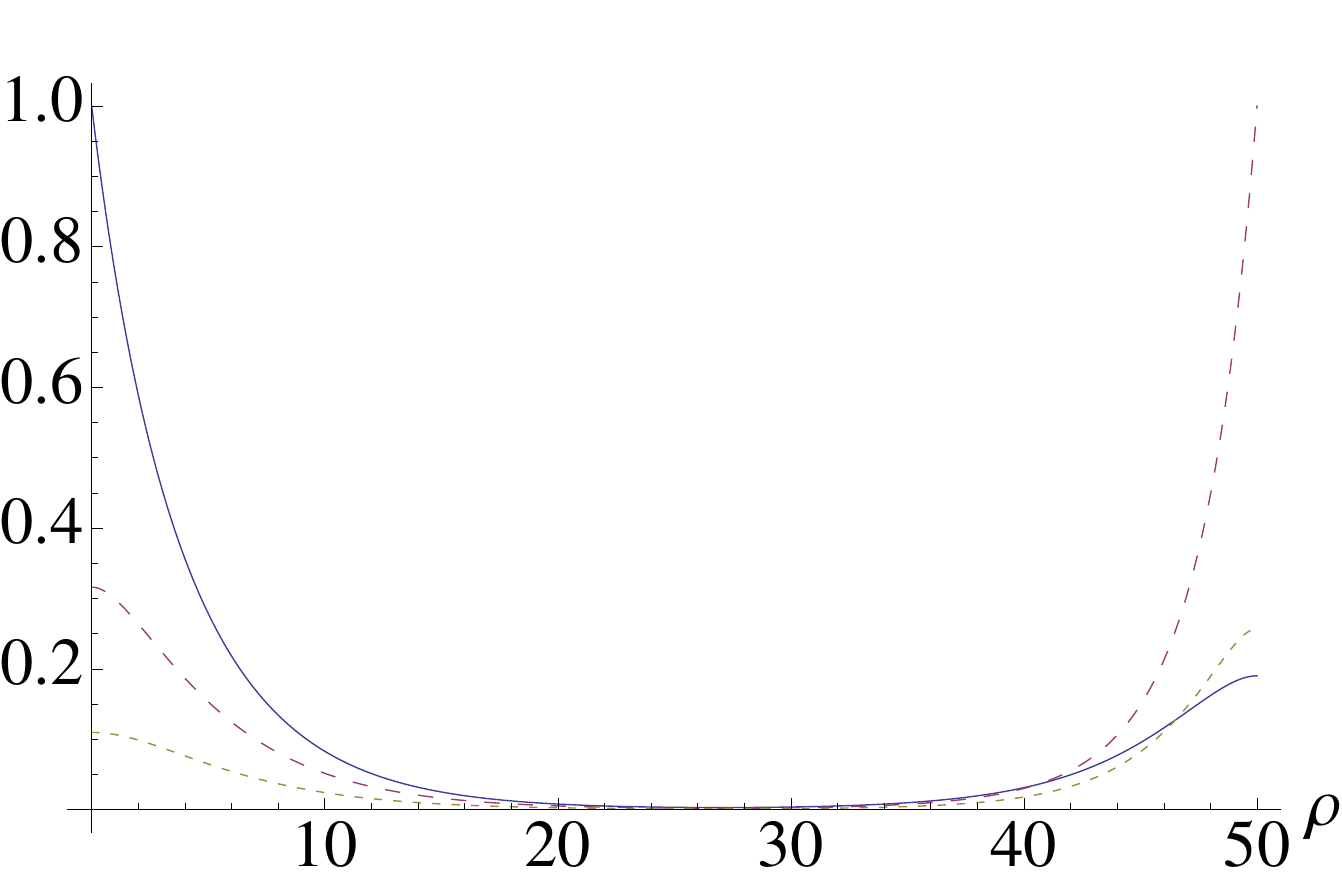} \caption{An $SU(4)$ two
monopole solution at $\tilde{R_{0}} = 5.9$. Thin dashed line is $g_{2}$,
medium dashed line is $g_{1}$ and solid line is $g_{0}$.}%
\label{figmix}%
\end{figure}

\section{Conclusions}

In the present paper, using the harmonic map formalism, we have extended the
investigation of BPS saturated t'Hooft-Polyakov monopoles in $\mathcal{M}%
^{2}\times S^{2}$ to the general case of $SU(N)$ gauge symmetry. We found
that, as per the $N=2$ case investigated previously in \cite{Canfora:2014jsa},
all equations for the monopole profile functions become autonomous. For some
specific cases we solved these equations numerically and in all cases we
demonstrated analytically the existence of non-trivial solutions in which the
field profiles are not proportional. Furthermore, we investigated the
remarkable power of the large $N$ limit in this scenario, where we have shown
that one can, under a suitable condition of smoothness, reduce the infinite set of profile equations to a single partial
differential equation for an interpolating function. This equation encodes the
subtle role played by the curvature parameter $\tilde{R}_{0}$ and elucidates
the flat $\tilde{R}_{0}\rightarrow\infty$ limit both from a perturbative and a
geometrical aspect. We leave the numerical treatment of this equation to
further work.\newline

\subsection*{Acknowledgements}

This work has been funded by the Fondecyt grants 1120352 and 3140122. The
Centro de Estudios Cient\'{\i}ficos (CECs) is funded by the Chilean Government
through the Centers of Excellence Base Financing Program of Conicyt.

\appendix{}

\section{Existence of solutions}
In this appendix we will provide a rigorous mathematical proof that solutions
to the set of equations obtained from eq.(\ref{set}) exist even in the case
that the fields are not proportional. This procedure does not yield the
solutions themselves, these are shown numerically in a previous section of the main text, but is
nonetheless an important part of the analysis of the field profile equations
and the space of their solutions. In particular, one of the inequalities
derived in this section will be useful in the discussion of the flat limit. \newline

The basic mathematical tool used here is the \textit{Schauder theorem} (see
for a detailed pedagogical review \cite{nonlinearanal}). For simplicity, we
will focus on the $SU(3)$ case but the same argument can be easily extended to
the general case.

The statement of the\textit{ Schauder theorem} (\cite{nonlinearanal}
\cite{nonlinearanal2}) is the following: let $S$ be a complete metric (Banach)
space so that a distance $d\left(  X,Y\right)  $ between any pair of elements
of the space is defined by
\begin{equation}
d\left(  X,Y\right)  \in%
\mathbb{R}
\ ,\ \ X,Y\in S\ , \label{ban1}%
\end{equation}
and such that, with respect to the chosen metric, from every Cauchy sequence
one can extract a convergent subsequence (this is the inclusion of ``complete"
in the definition of the metric space). Let $C$ be a bounded closed convex set
in $S$ and let $T$ be a compact operator\footnote{An operator $T$ from a
Banach space into itself (see, for a detailed discussion, \cite{nonlinearanal}
\cite{nonlinearanal2}) is called \textit{compact} if and only if, for any
bounded sequence $\left\{  x_{n}\right\}  $, the sequence $\left\{
T(x_{n})\right\}  $ has a convergent subsequence.} from the Banach space $S$
into itself such that $T$ maps $C$ into itself:%
\begin{equation}
T\left[  .\right]  :C\rightarrow C\ . \label{ban2}%
\end{equation}
Then the map $T\left[  .\right]  $ has (at least) one fixed point in $C$. In
other words, under the above hypothesis, there always exist a solution to the
equation%
\begin{subequations}
\begin{equation}
T\left[  X\right]  =X\ . \label{ban3}%
\end{equation}

Our task is to determine under what precise conditions this applies to the
system of equations obtained by decoupling the field profiles in
eq.(\ref{set}) for the case of $N=3$. To do so, let us rewrite the system in
eqs.(\ref{todasimilar1}) and (\ref{todasimilar2}) (with the factor $r$
replaced by $R_{0}$ everywhere) as coupled integral equations:%
\end{subequations}
\begin{equation}
g_{0}\left(  x\right)  =a_{0}+b_{0}x+\int_{0}^{x}\int_{0}^{s}\left[  \frac
{2}{\tilde{R_{0}}^{2}}\left\{  \exp\left[  g_{0}\left(  \rho\right)
-g_{1}\left(  \rho\right)  \right]  -1\right\}  +\frac{2}{\tilde{R_{0}}^{2}%
}\left[  \exp\left(  g_{1}\left(  \rho\right)  \right)  -1\right]  \right]
d\rho ds\ , \label{int1}%
\end{equation}%
\begin{equation}
g_{1}\left(  x\right)  =a_{1}+b_{1}x+\int_{0}^{x}\int_{0}^{s}\left[  \frac
{4}{\tilde{R_{0}}^{2}}\left[  \exp\left(  g_{1}\left(  \rho\right)  \right)
-1\right]  -\frac{2}{\tilde{R_{0}}^{2}}\left\{  \exp\left[  g_{0}\left(
\rho\right)  -g_{1}\left(  \rho\right)  \right]  -1\right\}  \right]  d\rho
ds\ , \label{int2}%
\end{equation}
where $a_{i}$ and $b_{i}$ represent the initial data for the two profiles
$g_{0}$ and $g_{1}$ and their derivatives at $x=0$. It is a trivial
computation to show that the above system of integral equations is equivalent
to the system in eq.(\ref{set}) with $N=3$. The system of eqs.(\ref{int1}) and
(\ref{int2}) can be written as a fixed point condition for the following
vectorial operator $\overrightarrow{T}$ acting on pairs of continuous
functions $\overrightarrow{g}\left(  x\right)  =\left(  g_{0}\left(  x\right)
,g_{1}\left(  x\right)  \right)  $:%
\begin{align*}
\overrightarrow{T}  &  :C^{0}\left[  0,L\right]  \times C^{0}\left[
0,L\right]  \rightarrow C^{0}\left[  0,L\right]  \times C^{0}\left[
0,L\right]  \ ,\\
\overrightarrow{g}\left(  x\right)   &  =\left(  g_{0}\left(  x\right)
,g_{1}\left(  x\right)  \right)  \in C^{0}\left[  0,L\right]  \times
C^{0}\left[  0,L\right]  \ ,
\end{align*}%
\begin{equation}
\overrightarrow{T}\left[  g_{0},g_{1}\right]  =\overrightarrow{T}\left[
\overrightarrow{g}\left(  x\right)  \right]  =\left(  T_{0}\left(  x\right)
,T_{1}\left(  x\right)  \right)  \ , \label{int2.5}%
\end{equation}
where
\begin{align}
T_{0}\left(  x\right)   &  =a_{0}+b_{0}x+\int_{0}^{x}\int_{0}^{s}\left[
\frac{2}{\tilde{R_{0}}^{2}}\left\{  \exp\left[  g_{0}\left(  \rho\right)
-g_{1}\left(  \rho\right)  \right]  -1\right\}  +\frac{2}{\tilde{R_{0}}^{2}%
}\left[  \exp\left(  g_{1}\left(  \rho\right)  \right)  -1\right]  \right]
d\rho ds\ ,\label{int3}\\
T_{1}\left(  x\right)   &  =a_{1}+b_{1}x+\int_{0}^{x}\int_{0}^{s}\left[
\frac{4}{\tilde{R_{0}}^{2}}\left[  \exp\left(  g_{1}\left(  \rho\right)
\right)  -1\right]  -\frac{2}{\tilde{R_{0}}^{2}}\left\{  \exp\left[
g_{0}\left(  \rho\right)  -g_{1}\left(  \rho\right)  \right]  -1\right\}
\right]  d\rho ds\ . \label{int4}%
\end{align}
The fixed-point condition is then simply%
\begin{equation}
\overrightarrow{g}\left(  x\right)  =\overrightarrow{T}\left[  \overrightarrow
{g}\left(  x\right)  \right]  \ , \label{fixedpoint}%
\end{equation}
where the operator $\overrightarrow{T}$ has been defined in eqs.
(\ref{int2.5}), (\ref{int3}) and (\ref{int4}). Whilst this proves that a fixed
point operator condition can be found, in order to apply Schauder's theorem we
have to show that $\overrightarrow{T}$ is a compact operator from a bounded
closed convex sub-set of a Banach space into itself. This is not a trivial
task, let us begin by defining the following metric in the space $C^{0}\left[
0,L\right]  \times C^{0}\left[  0,L\right]  $ (which is the Cartesian product
of the space of the continuous function on $\left[  0,L\right]  $\ with
itself):%
\begin{align}
d\left(  \overrightarrow{g},\overrightarrow{h}\right)   &  =\underset
{x\in\left[  0,L\right]  }{\sup}\left\vert g_{0}(x)-h_{0}(x)\right\vert
+\underset{x\in\left[  0,L\right]  }{\sup}\left\vert g_{1}(x)-h_{1}%
(x)\right\vert \ ,\label{ban4}\\
\overrightarrow{g}  &  =\left(  g_{0}\left(  x\right)  ,g_{1}\left(  x\right)
\right)  \ ,\ \ \overrightarrow{h}=\left(  h_{0}\left(  x\right)
,h_{1}\left(  x\right)  \right)  \ .\nonumber
\end{align}
With respect to this norm, the space $C^{0}\left[  0,L\right]  \times
C^{0}\left[  0,L\right]  $ is a Banach space (which we call $S$).

\indent Then we define a bounded closed convex sub-set $C$ of the Banach space
defined above (using the metric in eq.(\ref{ban4})) such that $\overrightarrow
{T}$ maps $C$ into itself,
\begin{equation}
C\equiv\left\{  \overrightarrow{g}\left(  x\right)  =\left(  g_{0}\left(
x\right)  ,g_{1}\left(  x\right)  \right)  \in\left.  S\right\vert \ \forall
x\in\left[  0,L\right]  \ \left\vert g_{0}(x)-a_{0}\right\vert \leq
B\ ,\ \ \left\vert g_{1}(x)-a_{1}\right\vert \leq B\right\}  \ ,\ \ B\in%
\mathbb{R}
_{+}\ , \label{ban5}%
\end{equation}
where $a_{i}$ are the initial data appearing in eqs.(\ref{int1}) and
(\ref{int2}) so that $C$ is closed by definition. It is easy to see that $C$
is bounded since%
\begin{align}
B  &  \geq\left\vert g_{0}(x)-a_{0}\right\vert \geq\left\vert g_{0}%
(x)\right\vert -\left\vert a_{0}\right\vert \Rightarrow\ \left\vert
g_{0}(x)\right\vert \leq B+\left\vert a_{0}\right\vert \ ,\label{ban6}\\
B  &  \geq\left\vert g_{1}(x)-a_{1}\right\vert \geq\left\vert g_{1}%
(x)\right\vert -\left\vert a_{1}\right\vert \Rightarrow\ \left\vert
g_{1}(x)\right\vert \leq B+\left\vert a_{1}\right\vert \ . \label{ban7}%
\end{align}
In order to prove that $C$ is convex we have to check that if $\overrightarrow
{g}\left(  x\right)  $ and $\overrightarrow{h}\left(  x\right)  $ both belong
to $C$ then $\theta\overrightarrow{g}\left(  x\right)  +\left(  1-\theta
\right)  \overrightarrow{h}\left(  x\right)  $ also belongs to $C$
$\forall\theta\in\left[  0,1\right]  $. This is easily verified as%
\begin{align}
\left\vert \theta g_{1}\left(  x\right)  +\left(  1-\theta\right)
h_{1}\left(  x\right)  -a_{1}\right\vert  &  \leq\left\vert \theta\left(
g_{1}\left(  x\right)  -a_{1}\right)  \right\vert +\left\vert \left(
1-\theta\right)  \left(  h_{1}\left(  x\right)  -a_{1}\right)  \right\vert
\leq\theta B+\left(  1-\theta\right)  B\leq B\ ,\label{ban8}\\
\left\vert \theta g_{0}\left(  x\right)  +\left(  1-\theta\right)
h_{0}\left(  x\right)  -a_{0}\right\vert  &  \leq\left\vert \theta\left(
g_{0}\left(  x\right)  -a_{0}\right)  \right\vert +\left\vert \left(
1-\theta\right)  \left(  h_{0}\left(  x\right)  -a_{0}\right)  \right\vert
\leq\theta B+\left(  1-\theta\right)  B\leq B\ . \label{ban9}%
\end{align}

Now we can proceed to show that $T$ is compact.\newline

First of all, we must show that if $\overrightarrow{g}_{n}\left(  x\right)
=\left(  g_{0}^{n}\left(  x\right)  ,g_{1}^{n}\left(  x\right)  \right)  $ is
a sequence in $C$ then the sequence $\overrightarrow{T}\left[  \overrightarrow
{g}_{n}\left(  x\right)  \right]  =\left(  T_{0}\left(  n;x\right)
,T_{1}\left(  n;x\right)  \right)  $ is uniformly bounded in $C$ (namely, the
absolute values of both components of $\overrightarrow{T}\left[
\ \overrightarrow{g}_{n}\left(  x\right)  \right]  $ are bounded by a constant
which does not depend on $n$ hence ensuring that $\overrightarrow{T}\left[
\ \overrightarrow{g}_{n}\left(  x\right)  \right]  $ belong to $C$ $\forall n$
as well). Therefore we consider%
\[
\left\vert T_{0}\left(  n;x\right)  \right\vert =\left\vert a_{0}+b_{0}%
x+\int_{0}^{x}\int_{0}^{s}\left[  \frac{2}{\tilde{R_{0}}^{2}}\left\{
\exp\left[  g_{0}^{n}\left(  \rho\right)  -g_{1}^{n}\left(  \rho\right)
\right]  -1\right\}  +\frac{2}{\tilde{R_{0}}^{2}}\left[  \exp\left(  g_{1}%
^{n}\left(  \rho\right)  \right)  -1\right]  \right]  d\rho ds\right\vert \leq
\]%
\[
\left\vert a_{0}\right\vert +\left\vert b_{0}L\right\vert +\int_{0}^{x}%
\int_{0}^{s}\left\vert \left[  \frac{2}{\tilde{R_{0}}^{2}}\left\{  \exp\left[
g_{0}^{n}\left(  \rho\right)  -g_{1}^{n}\left(  \rho\right)  \right]
-1\right\}  +\frac{2}{\tilde{R_{0}}^{2}}\left[  \exp\left(  g_{1}^{n}\left(
\rho\right)  \right)  -1\right]  \right]  \right\vert d\rho ds\Rightarrow
\]%
\begin{equation}
\left\vert T_{0}\left(  n;x\right)  \right\vert \leq\left\vert a_{0}%
\right\vert +\left\vert b_{0}L\right\vert +\frac{2L^{2}}{\tilde{R_{0}}^{2}%
}\left[  \left\{  \exp\left(  2B+\left\vert a_{0}\right\vert +\left\vert
a_{1}\right\vert \right)  -1\right\}  +\left[  \exp\left(  B+\left\vert
a_{1}\right\vert \right)  -1\right]  \right]  \ , \label{ban10}%
\end{equation}
and, similarly,
\begin{equation}
\left\vert T_{1}\left(  n;x\right)  \right\vert \leq\left\vert a_{1}%
\right\vert +\left\vert b_{1}L\right\vert +\frac{2L^{2}}{\tilde{R_{0}}^{2}%
}\left[  2\left[  \exp\left(  B+\left\vert a_{1}\right\vert \right)
-1\right]  +\exp\left[  2B+\left\vert a_{0}\right\vert +\left\vert
a_{1}\right\vert \right]  -1\right]  \ . \label{ban11}%
\end{equation}
In order to derive eqs.(\ref{ban10}) and (\ref{ban11}) we used that, because
of eqs. (\ref{ban6}) and (\ref{ban7}) (which are equivalent to saying that the
$\overrightarrow{g}_{n}\left(  x\right)  $ belong to $C$ $\forall\ n$), one
has%
\begin{align}
\left\vert \frac{2}{\tilde{R_{0}}^{2}}\left\{  \exp\left[  g_{0}^{n}\left(
\rho\right)  -g_{1}^{n}\left(  \rho\right)  \right]  -1\right\}  +\frac
{2}{\tilde{R_{0}}^{2}}\left[  \exp\left(  g_{1}^{n}\left(  \rho\right)
\right)  -1\right]  \right\vert  &  \leq\nonumber\\
\frac{2}{\tilde{R_{0}}^{2}}\left[  \left\{  \exp\left(  2B+\left\vert
a_{0}\right\vert +\left\vert a_{1}\right\vert \right)  -1\right\}  +\left[
\exp\left(  B+\left\vert a_{1}\right\vert \right)  -1\right]  \right]  \ ,\\
\left\vert \frac{4}{\tilde{R_{0}}^{2}}\left[  \exp\left(  g_{1}^{n}\left(
\rho\right)  \right)  -1\right]  -\frac{2}{\tilde{R_{0}}^{2}}\left\{
\exp\left[  g_{0}^{n}\left(  \rho\right)  -g_{1}^{n}\left(  \rho\right)
\right]  -1\right\}  \right\vert  &  \leq\nonumber\\
\frac{2}{\tilde{R_{0}}^{2}}\left[  2\left[  \exp\left(  B+\left\vert
a_{1}\right\vert \right)  -1\right]  +\exp\left[  2B+\left\vert a_{0}%
\right\vert +\left\vert a_{1}\right\vert \right]  -1\right]  \ ,\ \ \forall
\ n.
\end{align}
Eqs. (\ref{ban10}) and (\ref{ban11}) show that, if $\overrightarrow{g}%
_{n}\left(  x\right)  =\left(  g_{0}^{n}\left(  x\right)  ,g_{1}^{n}\left(
x\right)  \right)  $ is a sequence in $C$, the sequence $\overrightarrow
{T}\left[  \ \overrightarrow{g}_{n}\left(  x\right)  \right]  $ is uniformly
bounded. Moreover, one has to require that the sequence of images
$\overrightarrow{T}\left[  \ \overrightarrow{g}_{n}\left(  x\right)  \right]
$ belongs to $C$ as well. As always happens (see \cite{nonlinearanal} and
\cite{nonlinearanal2}) this will give some constraints on the range on the
parameters $B$, $L$ and $\tilde{R_{0}}$. In order for the sequence of images
to belong to $C$ the following inequalities must be satisfied (as can be
easily seen by comparing eqs. (\ref{ban5}), (\ref{ban6}) and (\ref{ban7}) with
eqs. (\ref{ban10}) and (\ref{ban11})):%
\begin{align}
\left\vert b_{0}L\right\vert +\frac{2L^{2}}{\tilde{R_{0}}^{2}}\left[  \left\{
\exp\left(  2B+\left\vert a_{0}\right\vert +\left\vert a_{1}\right\vert
\right)  -1\right\}  +\left[  \exp\left(  B+\left\vert a_{1}\right\vert
\right)  -1\right]  \right]   &  \leq B\ ,\label{ban10.5}\\
\left\vert b_{1}L\right\vert +\frac{2L^{2}}{\tilde{R_{0}}^{2}}\left[  2\left[
\exp\left(  B+\left\vert a_{1}\right\vert \right)  -1\right]  +\exp\left[
2B+\left\vert a_{0}\right\vert +\left\vert a_{1}\right\vert \right]
-1\right]   &  \leq B\ . \label{ban11.5}%
\end{align}
These imply that in order for this theorem to work, the length $L$ of the
cylindrically-shaped region in which these non-Abelian BPS monopoles are
living cannot exceed the bounds defined in eqs. (\ref{ban10.5}) and
(\ref{ban11.5}). One cannot obtain a very large value for the allowed $L$ by
simply increasing $B$ since the left hand sides of eqs. (\ref{ban10.5}) and
(\ref{ban11.5}) increase faster than the right hand sides, but the situation
improves if $\tilde{R_{0}}^{2}$ is very large, namely in the flat limit (in
which case the exponentials containing $B$ are suppressed). What is important
however is that it is always possible the choose $B$, $L$ and $\tilde{R_{0}}$
in such a way that eqs. (\ref{ban10.5}) and (\ref{ban11.5}) are fulfilled.
\newline\indent The next step to prove that $T$ is compact is to show that if
$\overrightarrow{g}_{n}\left(  x\right)  =\left(  g_{0}^{n}\left(  x\right)
,g_{1}^{n}\left(  x\right)  \right)  $ is a sequence in $C$ then the sequence
$\overrightarrow{T}\left[  \ \overrightarrow{g}_{n}\left(  x\right)  \right]
$ is \textit{equicontinuous}\footnote{A sequence of functions $\left\{
f_{n}\right\}  $ is said to be equicontinuous if, given $\epsilon>0$,
$\exists\ \delta>0$ such that $\left\vert f_{n}\left(  x\right)  -f_{n}\left(
y\right)  \right\vert <\epsilon$ whenever $\left\vert x-y\right\vert <\delta$
\textit{and, moreover, }$\delta$\textit{ does not depend on} $n$ (otherwise
the sequence would be continuous but not equicontinuous: see
\cite{nonlinearanal}).}. To show this, we must evaluate, \textit{for a
generic} $n$, the absolute values of following differences:%
\begin{align}
\left\vert T_{0}\left(  n;x\right)  -T_{0}\left(  n;y\right)  \right\vert  &
=\left\vert b_{0}\left(  x-y\right)  +\int_{x}^{y}\int_{0}^{s}\left[  \frac
{2}{\tilde{R_{0}}^{2}}\left\{  \exp\left[  g_{0}^{n}\left(  \rho\right)
-g_{1}^{n}\left(  \rho\right)  \right]  -1\right\}  +\frac{2}{\tilde{R_{0}%
}^{2}}\left[  \exp\left(  g_{1}^{n}\left(  \rho\right)  \right)  -1\right]
\right]  d\rho ds\right\vert \ ,\label{ban12}\\
\left\vert T_{1}\left(  n;x\right)  -T_{1}\left(  n;y\right)  \right\vert  &
=\left\vert b_{1}\left(  x-y\right)  +\int_{x}^{y}\int_{0}^{s}\left[  \frac
{4}{\tilde{R_{0}}^{2}}\left[  \exp\left(  g_{1}^{n}\left(  \rho\right)
\right)  -1\right]  -\frac{2}{\tilde{R_{0}}^{2}}\left\{  \exp\left[  g_{0}%
^{n}\left(  \rho\right)  -g_{1}^{n}\left(  \rho\right)  \right]  -1\right\}
\right]  d\rho ds\right\vert \ , \label{ban13}%
\end{align}
where $0<x<y<L$. After some trivial manipulations (which use the fact that all
the functions $\overrightarrow{g}_{n}\left(  x\right)  $ belong to $C$ and
consequently eqs.(\ref{ban6}) and (\ref{ban7}) are satisfied) one arrives at%
\begin{align}
\left\vert T_{0}\left(  n;x\right)  -T_{0}\left(  n;y\right)  \right\vert  &
\leq\left\vert x-y\right\vert \left[  \left\vert b_{0}\right\vert +\frac
{2L}{\tilde{R_{0}}^{2}}\left[  \left\{  \exp\left(  2B+\left\vert
a_{0}\right\vert +\left\vert a_{1}\right\vert \right)  -1\right\}  +\left[
\exp\left(  B+\left\vert a_{1}\right\vert \right)  -1\right]  \right]
\right]  \ ,\label{ban14}\\
\left\vert T_{1}\left(  n;x\right)  -T_{1}\left(  n;y\right)  \right\vert  &
\leq\left\vert x-y\right\vert \left[  \left\vert b_{1}\right\vert +\frac
{2L}{\tilde{R_{0}}^{2}}\left[  2\left[  \exp\left(  B+\left\vert
a_{1}\right\vert \right)  -1\right]  +\exp\left[  2B+\left\vert a_{0}%
\right\vert +\left\vert a_{1}\right\vert \right]  -1\right]  \right]  \ ,
\label{ban15}%
\end{align}
Thus, given any $\epsilon>0$, we can choose%
\begin{equation}
\delta<\frac{\epsilon}{2\left\{  \left\vert b_{0}\right\vert +\left\vert
b_{1}\right\vert +\frac{2L}{\tilde{R_{0}}^{2}}\left[  2\left[  \exp\left(
B+\left\vert a_{1}\right\vert \right)  -1\right]  +\exp\left[  2B+\left\vert
a_{0}\right\vert +\left\vert a_{1}\right\vert \right]  -1\right]  \right\}
}\ , \label{ban16}%
\end{equation}
in such a way that \textit{both} the choice of $\delta$ in eq. . (\ref{ban16})
does not depend on $n$ and,
\begin{equation}
\left\vert x-y\right\vert <\delta\Rightarrow\left\vert T_{i}\left(
n;x\right)  -T_{i}\left(  n;y\right)  \right\vert \leq\frac{\epsilon}%
{2}\ ,\ \ \forall\ n,\ \forall\ i=0,1\ . \label{ban18}%
\end{equation}

In summary, eqs. (\ref{ban10}), (\ref{ban11}), (\ref{ban10.5}) and
(\ref{ban11.5}) show that, if $\overrightarrow{g}_{n}\left(  x\right)
=\left(  g_{0}^{n}\left(  x\right)  ,g_{1}^{n}\left(  x\right)  \right)  $ is
any sequence in $C$, then the sequence $\overrightarrow{T}\left[
\ \overrightarrow{g}_{n}\left(  x\right)  \right]  $ is uniformly bounded in
$C$. Subsequently, eqs. (\ref{ban14}), (\ref{ban15}), (\ref{ban16}) and
(\ref{ban18}) show that, if $\overrightarrow{g}_{n}\left(  x\right)  =\left(
g_{0}^{n}\left(  x\right)  ,g_{1}^{n}\left(  x\right)  \right)  $ is any
sequence in $C$, then the sequence $\overrightarrow{T}\left[
\ \overrightarrow{g}_{n}\left(  x\right)  \right]  $ is
\textit{equicontinuous}. Consequently, using the \textit{Ascoli-Arzela'
theorem} (see \cite{nonlinearanal}), from any sequence $\overrightarrow
{T}\left[  \ \overrightarrow{g}_{n}\left(  x\right)  \right]  $ one can
extract a convergent subsequence: this implies that the operator
$\overrightarrow{T}$ is a compact operator from a bounded closed convex set
into itself.\newline\indent Finally, the Schauder theorem ensures that
eq.(\ref{fixedpoint}) (which is equivalent to our original system) has at
least one solution. Moreover, it is always possible to choose appropriately
the initial data $a_{i}$ and $b_{i}$ in such a way that the two profiles are
not proportional. This concludes the proof on existence of solutions of the
monopole profile equations. Moreover, one can also show by a similar procedure
that the solutions are actually not just continuous but they also have
continuous first and second derivatives.\footnote{There are many standard ways
to prove this result (see \cite{nonlinearanal} and \cite{nonlinearanal2}).
However, the easiest way to argue that this is indeed the case is by observing
that one can take the double derivative of the fixed-point formula directly
since the profiles are continuous and the right hand side of the fixed point
condition is a double integral of a continuous bounded function.}%
\newline\indent The present rigorous argument can be easily extended to the
$SU(N)$ case with $N>3$. Besides the intrinsic mathematical elegance of the
fixed-point Schauder-type argument, the present procedure also discloses the
presence of the bounds in eqs.(\ref{ban10.5}) and (\ref{ban11.5}) on the
length $L$ of the tube-shaped region in which these non-Abelian BPS monopoles
are living. At the present stage of the analysis, it is not possible yet to
say whether such a bound is just a limitation of the method or it signals some
deeper physical limitation on the volume of the regions in which one
constrains these non-Abelian BPS monopoles to live. Understanding whether or
not such BPS monopoles can fit into very large cylindrically-shaped regions is
certainly a very interesting question on which we hope to come back in a
future investigation.

\end{document}